\documentclass[12pt]{article}

\usepackage{amsmath}
\usepackage{amssymb}
\usepackage{amsthm}
\usepackage{amsfonts}
\usepackage{fullpage}
\usepackage{xspace}
\usepackage{graphicx}
\usepackage{slashed}

\usepackage{mathrsfs}

\usepackage{dsfont}

\begin{document}

\begin{titlepage}

\begin{flushright}
\end{flushright}
\vskip 2.5cm

\begin{center}
{\Large \bf Hadronic Lorentz Violation in\\
Chiral Perturbation Theory}
\end{center}

\vspace{1ex}

\begin{center}
{\large Rasha Kamand,\footnote{{\tt kamand@email.sc.edu}}
Brett Altschul,\footnote{{\tt baltschu@physics.sc.edu}}
and Matthias R. Schindler\footnote{{\tt mschindl@mailbox.sc.edu}}}

\vspace{5mm}
{\sl Department of Physics and Astronomy} \\
{\sl University of South Carolina} \\
{\sl Columbia, SC 29208} \\
\end{center}

\vspace{2.5ex}

\medskip

\centerline {\bf Abstract}

\bigskip

Any possible Lorentz violation in the hadron sector must be tied to Lorentz
violation at the underlying quark level. The relationships between the theories
at these two levels are studied using chiral perturbation theory. Starting from
a two-flavor quark theory that includes dimension-four Lorentz-violation operators,
the effective Lagrangians are derived for both pions
and nucleons, with novel terms appearing in both sectors. Since the Lorentz
violation coefficients for nucleons and pions are all related to a single
set of underlying quark coefficients, it is possible to place approximate bounds on pion
Lorentz violation using only proton and neutron observations. The resulting
bounds on four pion parameters are at the $10^{-23}$ level, representing improvements
of ten orders of magnitude.

\bigskip

\end{titlepage}

\section{Introduction}
\label{sec:intro}

Over the last twenty years, there has been a tremendous surge of interest
in the possibility that the fundamental Lorentz and CPT symmetries might
actually be violated in nature. Although no such symmetry violations have
yet been discovered experimentally, they might be part of a fundamental
theory of quantum gravity. In fact, if a violation of Lorentz symmetry is
ever discovered, it will be proof of new fundamental physics beyond the
standard model and general relativity, and the discovery will provide
important information about the nature of the new physics.

It has been recognized for a long time that these factors make precise
tests of Lorentz symmetry very important. More recently, however,
interest in Lorentz symmetry violation has grown because of the
development of an effective field theory, known as the standard model
extension (SME), that can be used to describe all forms of Lorentz
violation that may exist in a quantum field theory built around the
standard model fields~\cite{ref-kost1,ref-kost2}. The SME is quite general, and even
the minimal SME (mSME)---which is restricted to contain only gauge
invariant, superficially renormalizable operators in its action---includes
many more forms of Lorentz violation than previous studies had ever looked
at. The ability of the SME to parameterize such a wide array of
Lorentz-violating phenomena has led to a tremendous expansion in experimental
tests of Lorentz symmetry---in practically all sectors of the theory. An
up-to-date summary of the results of these tests may be found in~\cite{ref-tables}.

However, understanding of the SME is still far from complete. The SME is
formulated as a relativistic field theory, in terms of the fundamental
quark, lepton, gauge, and Higgs fields of the standard model, and the
relationships between the parameters in the fundamental SME Lagrangian and
experimental observables can be complicated. The most important
outstanding issue in this area arises from the fact that at low
energies, the standard model's strongly interacting degrees of freedom are
composite hadrons. There are many extremely precise constraints on
effective Lorentz violation coefficients for protons and neutrons, as well
as weaker constraints for other hadrons. However, it has not been possible
to translate these constraints into bounds on the more basic parameters
appearing in the SME action.

Using chiral perturbation theory ($\chi$PT) \cite{ref-weinberg,ref-gasser1,ref-gasser2} (also see
Ref.~\cite{Scherer:2012xha} for a pedagogical introduction), we shall examine the
relationships between quark- and hadron-level parameterizations of Lorentz
violation. (Another recent $\chi$PT analysis~\cite{ref-noordmans} has looked at a quite
different set of terms from the ones we shall be considering.)
In section~\ref{sec-LVquark}, we introduce Lorentz violation for the
fundamental quark fields. Then, in section~\ref{sec-LVhadron},
we construct the $\chi$PT action in two-flavor QCD for
both pions and nucleons, at leading order (LO) in each sector. This will reveal
which Lorentz-violating parameters in the quark sector contribute to
particular operators built out of the hadron fields. We shall
find that even at the lowest chiral orders, there are expected to be terms
in the hadronic theory that have previously not been studied. Using the
relationships we have uncovered between the actions in different sectors, we
shall look at how bounds on the Lorentz-violating behavior of one type
of hadron may be used to place constraints on different phenomena
involving entirely different species of particles in section~\ref{sec-exper}.
Finally, section~\ref{sec-concl} summarizes our main results and the outlook for
future work.

\section{Lorentz Violation at the Quark Level}
\label{sec-LVquark}

The basic idea behind the SME is to look at an effective field theory
containing operators constructed out of the standard model fields, but
which does not respect Lorentz symmetry. The novel operators can have
arbitrary free Lorentz indices. These indices are contracted with
tensor-valued coefficients, which serve as preferred vector or tensor
backgrounds in spacetime. Otherwise
identical experiments done in different reference frames
may yield different outcomes because of these background tensors. The
vector and tensor backgrounds may be constrained by comparing the results
of experiments with the apparatus in different orientations relative to
the fixed stars, or experimental setups moving with different velocities.

There are now many strong constraints on the Lorentz-violating parameters
of the mSME, coming from experiments in atomic, nuclear, and astroparticle
physics. Any local, stable field theory that violates CPT symmetry also
violates Lorentz invariance~\cite{ref-greenberg}. So the SME is also the unique well-behaved
effective field theory governing CPT violation, and the mSME is
also used for parameterizing constraints on CPT violation. In most (but
not all) cases, Lorentz-violating operators with odd numbers of
free Lorentz indices are CPT odd, while those with even number of indices
are CPT even. In this paper, we shall only be considering CPT-even
operators with two free Lorentz indices.

We are also restricting our attention to the quark sector of the
mSME, so the possible Lorentz-violating operators will be built out of quark
field bilinears. These are the fundamental operators of the effective field
theory. However, it is conventional when studying hadronic systems to
consider similar Lorentz-violating operators for the composite quanta---protons,
neutrons, pions, and such. Perhaps the most important basic question that
remains about the structure of the SME is the problem of relating the operators
at the hadron level to quark-level operators and thus translating constraints
resulting from experiments performed on real hadrons to bounds on the underlying
quark parameters. We will use $\chi$PT to begin addressing this question.

Of course, there could also be Lorentz violation in the gluon sector of the
mSME. There are operators in the pure $SU(3)_{c}$ gauge sector with the same
symmetries as the coefficients we shall be considering in this paper. They
could contribute to some of the same hadronic coefficients as the quark 
coefficients listed below. So these gauge-sector parameters may be an important
object of future study.

Moreover, there are also additional quark operators, besides those considered in this paper. We shall
be looking solely at operators with mass dimension 4. These operators,
because their Lorentz structures are similar to the structure of the conventional
kinetic term for Dirac fermions, give rise to a rather complicated set of
phenomena. Mixing between the conventional kinetic term and the
dimension-4 Lorentz violation can give rise to a number of effects that have
no analogues in models with other kinds of Lorentz-violating operators.
So Lorentz-violating operators of mass dimension 3, although we will not be
considering them in this paper, are expected to have a more limited array of
physical effects; analyzing them will be the subject of future work.

This leaves the portion of the SME Lagrange density relevant to our present
work, given by~\cite{ref-kost2}
\begin{equation}
\label{qrk_cpt_even}
\mathcal{L}_\text{quark}^\text{CPT-even}=i(c_{Q})_{\mu\nu AB}\bar{Q}_{A}\gamma^{\mu}
D^{\nu}Q_{B}+i(c_{U})_{\mu\nu AB}\bar{U}_{A}\gamma^{\mu}D^{\nu}U_{B}
+ \,i(c_{D})_{\mu\nu AB}\bar{D}_{A}\gamma^{\mu}D^{\nu}D_{B}.
\end{equation}
The covariant derivatives contain the standard model gauge fields, and in curved
spacetime the derivatives must be taken as linear combinations of derivative
operators acting to the right and left.
The left- and right-handed quark multiplets are denoted by
\begin{equation}
Q_{A}= 
\left[\begin{array}{c}
u_{A} \\
d_{A}
\end{array}
\right]_{L}
\quad U_{A} = \left[u_{A}\right]_{R}
\quad D_{A} = \left[d_{A}\right]_{R}, 
\end{equation}
and $A,B = 1,2,3$ label the quark generations.
The $c_{\mu\nu}$ parameters in eq.~\eqref{qrk_cpt_even} are dimensionless coupling coefficients that are
Hermitian in
the quark generation space spanned by $A$ and $B$, while $\mu$ and $\nu$ are spacetime indices.
Restricting the Lagrange density of Eq.~\eqref{qrk_cpt_even} to up ($u$) and down ($d$) quarks, we may rewrite it
as\footnote{For a more complete analysis, one should explicitly integrate out the heavy degrees of freedom via the
renormalization group.}
\begin{equation}
\label{re-written_lag}
\mathcal{L}_\text{light quarks}^{\text{CPT-even}} 
=i\bar{Q}_{L} C_{L\mu\nu} \gamma^{\mu} D^{\nu} Q_{L}
+ i\bar{Q}_{R} C_{R\mu\nu}  \gamma^{\mu} D^{\nu} Q_{R},
\end{equation}
where now $Q_{L/R}=[u_{L/R},d_{L/R}]^T$ and the couplings are collected in the matrices
\begin{equation}
C_{L/R}^{\mu\nu}=\left[
\begin{array}{cc}
c^{\mu\nu}_{u_{L/R}} & 0 \\
0 & c^{\mu\nu}_{d_{L/R}}
\end{array}
\right].
\end{equation}

Note that this formalism allows for there to be different $c^{\mu\nu}$ coefficients for
the left-handed $u$ and $d$ quarks. Physically, the $SU(2)_{L}$ gauge invariance
of the mSME requires that the coefficients for these two chiral fermion species
be identical. Moreover, the separate coefficients for left- and right-handed chiral fermions
are not typically what are observed in experiments with baryons. Experimental constraints
are typically placed on the combinations $c^{\mu\nu}=\frac{1}{2}(c_{L}^{\mu\nu}+c_{R}^{\mu\nu})$
and $d^{\mu\nu}=\frac{1}{2}(c_{L}^{\mu\nu}-c_{R}^{\mu\nu})$.

It may also be convenient to split the coefficients into isosinglet and
isotriplet pieces.  These are $^{1}C^{\mu\nu}_{L/R}=\mathrm{Tr}
(C_{L/R}^{\mu\nu})$ and $^{3}C^{\mu\nu}_{L/R}=C^{\mu\nu}_{L/R}-(\mathds{1}/2)\,
^{1}C^{\mu\nu}_{L/R}$ (where $\mathds{1}$ is the identity in flavor space).

It will frequently be important that the portions of these Lorentz-violating two-index
tensors that are antisymmetric in their Lorentz indices cannot be observed at linear order.
Only at second order in the Lorentz violation do these antisymmetric combinations have physical
effects. This is a consequence of the fact that field redefinitions such as
$Q_{L}'=[1-(i/2)(C_{L}^{\mu\nu}-C_{L}^{\nu\mu})\sigma_{\mu\nu}]Q_{L}$ can actually eliminate
the antisymmetric terms from the Lagrange density at first order~\cite{ref-colladay2}.
In addition, as discussed below, in the absence of external gauge fields the antisymmetric terms do not
contribute to the effective
hadronic Lagrange density at leading order in the chiral power counting. We will thus assume $C_{L/R}^{\mu\nu}$
to be symmetric in
the following.

Naively constructed Lagrangian terms involving the antisymmetric parts of the $C_{L/R}^{\mu\nu}$ tensors would
also be
odd under charge conjugation---depending on such left-right differences as
$c^{\mu\nu}_{u_{L}}-c^{\mu\nu}_{u_{R}}$. Such C-odd terms cannot exist in the purely hadronic sector; they
require the existence of additional external fields, which we are not considering here. However, the existence of
C-odd bosonic terms that
are antisymmetric in Lorentz indices and do explicitly involve external gauge fields was previously noted
in~\cite{ref-altschul30}.
We suspect that they may be a ubiquitous feature of Lorentz-violating theories with spin-0 excitations and
external fields. However, very little is understood about these inherently Lorentz-violating terms.

Many of the best experimental bounds on Lorentz violation for nucleons
come from atomic clock experiments involving nuclear magnetic transitions, and these
are typically analyzed using only effective proton and neutron $c^{\mu\nu}$ and $d^{\mu\nu}$
coefficients. Yet there should also be additional couplings (analogous to the
hadrons' anomalous magnetic moments in the usual Lorentz-invariant theory) that appear in
the Lorentz-violating effective action for hadrons interacting with an external
electromagnetic field. The omission of these terms from existing analyses means that many
quoted bounds should really only be interpreted as order of magnitude constraints. While we
shall not consider the hadronic sector coupled to external fields in this paper, our
$\chi$PT methodology may easily be adapted to study such terms in the future.

\section{Lorentz-Violating Hadronic Lagrangian}
\label{sec-LVhadron}

With the quark-level Lagrange density established, we are in a position to construct the
effective Lagrangian at the hadronic level. This is done by
considering all terms allowed by the symmetries of the underlying
theory~\cite{ref-weinberg,ref-gasser1,ref-gasser2}.
For quantum chromodynamics (QCD), these symmetries include the discrete operations C, P, and T. 
In addition, QCD possesses an accidental chiral symmetry in the limit of vanishing quark masses. 
Physically, the masses of the $u$ and $d$ quarks are much smaller than the masses of typical hadrons,
which means that setting $m_u=m_d = 0$ is a reasonable starting point for the construction of the effective
Lagrangian. Further it is known that the resulting $SU(2)_L \times SU(2)_R$ symmetry is spontaneously broken to
the diagonal group $SU(2)_V$,
and the pions are considered the associated Goldstone bosons. 
The corresponding pion fields are collected in the $SU(2)$ matrix~\cite{ref-coleman1}
\begin{equation} 
U(x) = \exp\left[ i \frac{\phi(x)}{F} \right],
\end{equation}
where $\phi = \sum \phi_a\tau_a$ in terms of the $SU(2)$ generators,
and $F\approx 92.4\,\text{MeV}$ is the pion decay constant in the $SU(2)$ chiral limit. 
Under chiral transformations $U(x)$ transforms according to
\begin{equation}
U(x) \rightarrow U'(x) = RU(x)L^{\dagger},
\end{equation}
where $(L,R) \in SU(2)_{L} \times SU(2)_{R}$.
The mesonic effective Lagrangian in the chiral limit without the coupling to external fields can then be
constructed in terms of $U(x)$ and its derivatives.
At low energies the $\chi$PT power counting dictates that derivatives acting on the pion fields
are suppressed, and the effective Lagrangian can be organized in terms of the number of derivatives.
The leading order term is given by\footnote{We use $\mathscr{L}$ to denote Lorentz-conserving and $\mathcal{L}$ 
for Lorentz-violating Lagrange densities.}
\begin{equation}
\label{lo-chpt}
\mathscr{L}_{\pi}^\text{LO}=\frac{F^2}{4} \text{Tr}(\partial_\mu U \partial^{\mu} U^\dagger),
\end{equation}
with the trace Tr being taken over flavor space.

In addition to spontaneous symmetry breaking, chiral symmetry is also explicitly broken by the non-zero quark
masses. The mass terms for the light $u$ and $d$ quarks may be written as 
\begin{equation}
\mathscr{L}_{\mathcal{M}} = -\bar{Q}_{R}\mathcal{M}Q_{L} - \bar{Q}_{L}\mathcal{M}^{\dagger}Q_{R}
\end{equation}
with the quark mass matrix $\mathcal{M} = \text{diag}[m_u,m_d]$. 
Under chiral transformations of the right- and left-handed quark fields, $Q_{R}\to RQ_{R}$
and $Q_{L}\to LQ_{L}$, the mass term is not invariant. 
However, the pattern of symmetry breaking can be matched onto the effective chiral Lagrangian by assuming that
the mass matrix transforms as $\mathcal{M} \to R\mathcal{M}L^\dagger$.
The lowest-order chirally invariant term that is also even under C and P is then given by\footnote{The subscript
``s.b.''~refers to symmetry breaking.}
\begin{equation}
\mathscr{L}_\text{s.b.}^\text{LO} = \text{Tr}(\mathcal{M} U^\dagger + U\mathcal{M}^\dagger).
\end{equation}
This term contributes at the same chiral order as the term in eq.~\eqref{lo-chpt}.

Nucleon fields can also be considered. 
Under chiral transformations the nucleon doublet $\Psi=[p,n]^T$ transforms
as~\cite{ref-coleman1,ref-callan1,ref-georgi}
\begin{equation}
\Psi \to K(L,R,U)\Psi \ .
\end{equation}
The $SU(2)$-valued function $K(L,R,U)$ is defined by
\begin{equation}
u(x) \to u'(x) = \sqrt{RUL^{\dagger}} \equiv RuK^\dagger(L,R,U) = KuL^{\dagger}\ ,
\end{equation}
where $[u(x)]^{2} = U(x)$.
Because $K$ depends on the pion fields through $U(x)$, the covariant derivative of the nucleon field,
\begin{equation}
D_{\mu}\Psi = (\partial_{\mu} + \Gamma_{\mu})\Psi,
\end{equation} 
contains pion fields in the chiral connection $\Gamma_\mu$~\cite{Gasser:1987rb}.
Recall that we have neglected the coupling of the nucleon field to external gauge fields, so the
only term in the connection is
\begin{equation}
\Gamma_{\mu} = \frac{1}{2}(u^{\dagger}\partial_{\mu}u + u\partial_{\mu}u^{\dagger}).
\end{equation}
The lowest-order pion-nucleon Lagrange density takes the form
\begin{equation}
\mathscr{L}_{\pi N}^\text{LO}=
\bar{\Psi} \left( \slashed{D} - m +\frac{g_A}{2} \gamma^\mu\gamma_5 u_\mu \right) \Psi,
\end{equation}
where $m$ is the nucleon mass, and $g_A$ the axial vector coupling in the chiral limit. In the absence of
external gauge fields
\begin{equation}
u_\mu = i(u^{\dagger}\partial_{\mu}u - u\partial_{\mu}u^{\dagger}).
\end{equation}

In order to construct the effective Lagrangian including Lorentz violation in terms of hadronic degrees of
freedom, we have to match symmetry properties of the quark-level expression from eq.~\eqref{re-written_lag} onto
the hadronic level. In particular, under chiral transformations, $Q_{R}\to RQ_{R}$, $Q_{L}\to LQ_{L}$, the
Lagrange density of eq.~\eqref{re-written_lag} transforms as
\begin{equation}
\mathcal{L}_\text{light quarks}^{\text{CPT-even}} \to
i\bar{Q}_{L} L^{\dagger} C_{L\mu\nu} L \gamma^{\mu} D^{\nu} Q_{L}
+ i\bar{Q}_{R} R^{\dagger} C_{R\mu\nu} R \gamma^{\mu} D^{\nu} Q_{R}.
\end{equation}
The matrices $C_{L/R}^{\mu\nu}$ are constant, and chiral symmetry is broken by the terms in
eq.~\eqref{re-written_lag}. Following the method described above for the quark mass terms in the QCD Lagrangian,
we note that the Lorentz-violating
action {\em would be} invariant under chiral transformations {\em if} $C_{L/R}^{\mu\nu}$ transformed as
\begin{equation}
\label{eq:CLRtrans}
C_L^{\mu\nu}   \to L C_L^{\mu\nu} L^{\dagger},\quad 
C_R^{\mu\nu}  \to R C_R^{\mu\nu} R^{\dagger}.
\end{equation}
Because of the cyclic property of the trace, this implies for the isosinglet and isotriplet components
\begin{align}
\label{eq:isoCLRtrans}
&^{1}C_L^{\mu\nu}   \to {^{1}C_L^{\mu\nu}} ,\quad {^{3}C_L^{\mu\nu}}   \to L{^{3} C_L^{\mu\nu}} L^{\dagger},
\quad \\
&^{1}C_R^{\mu\nu}  \to {^{1}C_R^{\mu\nu}} , \quad  {^{3}C_R^{\mu\nu}}  \to R{^{3}C_R^{\mu\nu}} R^{\dagger}.
\nonumber
\end{align}
Using this transformation behavior to construct a Lagrange density that is invariant under chiral transformations,
the pattern of symmetry breaking in the quark-level action can be matched onto the hadronic Lagrangian. 

With these basic building blocks---and assuming the transformation behavior given by eq.~\eqref{eq:CLRtrans}---we
may construct the chirally invariant, Lorentz-violating leading order effective Lagrange densities for the
pure pion sector and for pion-nucleon interactions.
Writing down all possible expressions satisfying the symmetry properties produces some terms that are linearly
dependent.
By applying integration by parts and the symmetry properties of the $C_{\mu\nu}$ tensors, the number of terms may
be reduced. Moreover, the transformation properties under parity and charge conjugation will also produce
relationships among
the various terms. 
The Lorentz-violating terms in the quark-level Lagrange density are the only potential sources of $C$, $P$, and
$T$ violations in this theory. So at leading order, the terms in the pion Lagrange density need to
have the same discrete symmetries as the terms in the underlying quark density that are multiplied
by the same $C_{L/R}^{\mu\nu}$ coefficients. This forces the coefficients for left- and right-handed
quark fields to enter the pion Lagrange density multiplied by the same numerical low-energy couplings (LECs),
drastically reducing the number of independent terms.

The LO minimal mesonic Lagrange density is given by
\begin{eqnarray}
\label{lo_pionlagrangian}
\mathcal{L}_{\pi}^{\text{LO}} & = &
\beta^{(1)}\frac{F^2}{4}\left(^{1}C_{R\mu\nu} +\,^{1}C_{L\mu\nu}\right)
\mathrm{Tr}[(\partial^{\mu}U)^{\dagger} \partial^{\nu}U] , 
\end{eqnarray}
where $\beta^{(1)}$ is a dimensionless LEC. It encodes short-distance physics and cannot be determined
from symmetry arguments. In principle, it could be calculated using nonperturbative QCD; however, the
required calculations are not currently available.
The factor of $F^2/4$ is present to mirror the form of the standard pion Lagrange density and is also chosen such
that based on naive dimensional
analysis~\cite{Manohar:1983md} $\beta^{(1)}$ is expected to be of natural size, i.e. $\mathcal{O}(1)$.

Chiral symmetry also allows an analogous term for the isotriplet components $^{3}C_{R/L}^{\mu\nu}$ with an
independent LEC, 
\begin{equation}
\label{lo_pion_isotriplet}
\beta^{(2)}\frac{F^2}{4} \mathrm{Tr}[(\partial^{\mu}U)^{\dagger}\, ^{3}C_{R\mu\nu} \partial^{\nu}U +
\partial^{\mu}U\, ^{3}C_{L\mu\nu} (\partial^{\nu}U)^{\dagger}].
\end{equation}
However, this term can be shown to vanish because of the symmetry of $^{3}C_{R/L}^{\mu\nu}$ in the Lorentz indices.

In principle there is a second, nearly-identical-looking copy of the Lagrange density of eqs.~\eqref{lo_pionlagrangian} and \eqref{lo_pion_isotriplet} contracted with the
antisymmetric parts of the $C_{L/R}^{\mu\nu}$. These terms would be accompanied by an independent set of LECs. However, all the
terms involved can be shown to be total derivatives, so they
may be dropped; and thus only the symmetric part of the $C_{L/R}^{\mu\nu}$ contributes at leading order.

Expanding $U(x)$ in terms of the pion fields shows that the Lagrange density in eq.~\eqref{lo_pionlagrangian}
not only contains corrections to the pion propagator, but also induces new multi-pion interactions.
The two-pion portion of the Lagrange density is
\begin{equation}
\label{lo_2pionlagrangian}
\mathcal{L}_{\pi}^{\text{LO},2\phi}=\frac{\beta^{(1)}}{2}(c^{\mu\nu}_{u_{L}}+c^{\mu\nu}_{d_{L}}+
c^{\mu\nu}_{u_{R}}+c^{\mu\nu}_{d_{L}})\partial_{\mu}\phi_{a}\partial_{\nu}\phi_{a}.
\end{equation}
We shall defer most discussion of the term involving just two pion field operators until section~\ref{sec-exper},
because such terms lead to the propagator modifications that have been used to constrain Lorentz violation
in the pion sector.

For the moment, we shall concentrate on the forms taken by the pion vertices.
Unfortunately, all three-pion vertices vanish
when the symmetric parts of the $c_{L/R}^{\mu\nu}$ are involved.
The four-pion vertex takes the form
\begin{eqnarray}
\label{eq-lo4pi}
\mathcal{L}_{\pi}^{\text{LO},4\phi} & = &
\frac{\beta^{(1)}}{6F^{2}}(c^{\mu\nu}_{u_{L}}+c^{\mu\nu}_{d_{L}}+
c^{\mu\nu}_{u_{R}}+c^{\mu\nu}_{d_{L}})(\phi_{a}\phi_{b}\partial_{\mu}\phi_{a}\partial_{\nu}\phi_{b}
-\phi_{b}\phi_{b}\partial_{\mu}\phi_{a}\partial_{\nu}\phi_{a}).
\end{eqnarray}
This term is a straightforward Lorentz-violating generalization of the
usual four-pion vertex. 
Many Lorentz-violating operators in the SME Lagrange density are structurally similar to operators found in the usual
standard model. For example,
the quark kinetic terms from eq.~\eqref{re-written_lag} resemble standard kinetic terms, but instead
of the indices on $\gamma^{\mu}$ and $D^{\nu}$ being contracted with the metric tensor $g_{\mu\nu}$,
they are contracted with the Lorentz-violating backgrounds. The four-pion vertex can be similarly viewed as a deformation of the standard model four-pion vertex. 
Vertices with more pion fields can similarly be derived.

Looking at the pion sector overall,
the two-pion term includes a $k^{\mu\nu}$-type term that modifies the pion propagation. Terms of
this type have previously been studied and experimentally constrained, and we shall discuss the
physics of such terms in further detail in section~\ref{sec-exper}. The Lorentz-violating multi-pion
interaction terms are new and have never been written down before, although the four-pion terms have
a relatively straightforward structure that could probably be guessed at fairly easily.

However, it
is an important observation that this vertex involves exactly the same $c^{\mu\nu}_{u}$ and
$c^{\mu\nu}_{d}$ parameters that appear in the pion propagation Lagrangian. In the Lorentz-violating
effective field theory approach, the Lorentz violation often has to be described using
completely separate and independent coefficients multiplying the various operators. Understanding the
relationships among these coefficients requires additional information---either about the physics
underlying the Lorentz violation, or about the symmetry properties of the
low-energy theory. For example, $U(1)$
gauge invariance can ensure that the same Lorentz violation coefficients appear in the kinetic
term for a charged species and in the coupling to the electromagnetic field. In the pion
theory, the chiral symmetry of the underlying physics provides concrete relationships between
Lorentz-violating operators involving different numbers of fields. Since the two- and four-pion
operators also involve the same LEC $\beta^{(1)}$, it would thus be possible to place
constraints on purely interactional effects by looking at free
particle propagation phenomena, and vice versa.

There are also relationships between Lorentz-violating behavior in the pure pion sector and Lorentz
violation for
baryons---for nucleons, in particular. The minimal LO baryonic Lagrange density is (recalling that
$\Psi$ is the nucleon doublet field)
\begin{eqnarray}
\label{lo_pionucleonlagr}
\mathcal{L}_{\pi N}^{\text{LO}} & = & \Big\{
\alpha^{(1)}\bar{\Psi}[(u^{\dagger}\;{^3C_{R}^{\mu\nu}} u + u\: {^3C_{L}^{\mu\nu}} u^{\dagger})(\gamma_{\nu} i D_{\mu} +
\gamma_{\mu} i D_{\nu})]\Psi \\
& & +\,\alpha^{(2)}\left({^1C_{R}^{\mu\nu}} + \: {^1C_{L}^{\mu\nu}}\right)\bar{\Psi}(\gamma_{\nu} i D_{\mu} +
\gamma_{\mu}i D_{\nu})]\Psi \nonumber\\
& & +\,\alpha^{(3)}\bar{\Psi}[(u^{\dagger}\;{^3C_{R}^{\mu\nu}} u - u\: {^3C_{L}^{\mu\nu}} u^{\dagger})(\gamma_{\nu}\gamma^{5}
i D_{\mu} + \gamma_{\mu}\gamma^{5} i D_{\nu})]\Psi \nonumber\\
& & +\,\alpha^{(4)}\left({^1C_{R}^{\mu\nu}} - \: {^1C_{L}^{\mu\nu}}\right)\bar{\Psi}(\gamma_{\nu}\gamma^{5}
i D_{\mu} + \gamma_{\mu}\gamma^{5} i D_{\nu})\Psi\Big\}, \nonumber
\end{eqnarray}
where the $\alpha^{(n)}$'s are dimensionless LECs that by naive dimensional analysis are expected to be $\mathcal{O}(1)$.

These four operators exhaust the possibilities at this order. Recall that these operators were formed by writing down all
combinations of nucleon operators that would be chirally invariant if the Lorentz violation tensors transformed
according to eq.~\eqref{eq:isoCLRtrans}. This essentially requires that the isotriplet components of the Lorentz violation tensors
be sandwiched between $u$ and $u^{\dag}$ to give $u^{\dagger}C_{R}^{\mu\nu}u$ and $uC_{L}^{\mu\nu} u^{\dagger}$. 

There must also be two free Lorentz indices to be contracted with the $C^{\mu\nu}_{L/R}$
background tensors; and since we know
in advance that only the operators that are symmetric in their Lorentz indices will contribute at leading order, it
is advantageous to take the $C_{L/R}^{\mu\nu}$ to be symmetric from the start. This avoids the presence of terms
such as those containing $[D_{\mu},D_{\nu}]$. In addition, the antisymmetric combination of two nucleon covariant derivatives is
of higher order in the $\chi$PT power counting as well \cite{ref-fettes}.
Note that $\mathcal{L}^{\text{LO}}_{\pi N}$ contains operators with the same two kinds of
structures in spinor space as the underlying quark Lagrange density. These are, of course, not the only structures
with two free Lorentz indices that may be
constructed out of Dirac matrices and covariant derivatives. For example, the operator could contain additional
$D^{\mu}D_{\mu}$ or $\gamma^{\mu}D_{\mu}$ terms sandwiched between $\bar{\Psi}$ and $\Psi$. However, these terms can
be eliminated using the equations of motion. For example, $i\gamma^{\mu}D_{\mu}\Psi=m_{N}\Psi$ (where $m_{N}$ is the nucleon
mass), up to higher-order chiral corrections; so at leading order, any term with an additional $\gamma^{\mu}D_{\mu}$
may be absorbed into one of the terms given in eq.~\eqref{lo_pionucleonlagr}. A dictionary of the possible reductions
is given in~\cite{ref-fettes}.

There are really eight terms in eq.~\eqref{lo_pionucleonlagr}, arranged in pairs. As in the pion case, the discrete
symmetries of the underlying quark theory force there to be specific relationships between the operators involving
$C_{L}^{\mu\nu}$ and those with $C_{R}^{\mu\nu}$. There is no C violation in the chiral dynamics, so fermion operators
that are even under C must be multiplied by likewise C-even combinations of $C_{L/R}^{\mu\nu}$ coefficients; this means
symmetric sums of the corresponding coefficients for right- and left-handed quarks. Conversely, any C violation in the
theory must be generated by C violation in the pattern of $C_{L/R}^{\mu\nu}$ coefficients. So a C-odd fermion operator
must be multiplied by a difference between right- and left-chiral Lorentz violation coefficients. The well known
transformation properties of Dirac bilinears indicate that the terms involving $\gamma_{5}$ are the ones that are odd
under C. This accounts for the extra negative signs in the $\alpha^{(3)}$ and $\alpha^{(4)}$ terms; while eq.~\eqref{re-written_lag}
with $C_{L}^{\mu\nu}=C_{R}^{\mu\nu}$ is even under C, with $C_{L}^{\mu\nu}=-C_{R}^{\mu\nu}$ it is odd.

As in the pion sector, there are two distinct ways that the Lorentz-violating tensors may enter. They may be
traced over the flavor space, outside the Dirac bilinear, or they may be inserted between the two-flavor baryon spinor
$\Psi$ and its adjoint $\bar{\Psi}$. These two possibilities
give different kinds of contributions to the separate Lorentz violation
coefficients for protons and neutrons. The terms with traces sum over the $u$ and $d$ quark terms uniformly, giving the
same contributions to the baryon Lorentz violation coefficients, regardless of isospin. However the terms with
the $C_{L/R}^{\mu\nu}$ tensors actually contained within the fermion bilinear give quite different coefficients for the
two nucleon species.

Finally, we point out that,
because of the presence of the covariant derivatives in eq.~\eqref{lo_pionucleonlagr}, this
nucleon Lagrange density actually contains interaction terms with arbitrary numbers of pions. These may be expanded
directly, although we shall not consider them in further detail, because they are not presently useful for placing
constraints on any Lorentz violation coefficients.

\section{Experimental Constraints}
\label{sec-exper}

We shall now turn our attention to setting new constraints on the effective Lorentz violation
coefficients in one hadronic sector using experimental observations made in an entirely different
sector. The key point is that there are only a limited number of underlying Lorentz violations for
the quarks, which determine the effective coefficients for a much larger number of meson and baryon types.
It is possible to measure or bound some combination of the quark coefficients using one type of hadron and
transfer this information to another kind of particle entirely. The presence of numerous LECs limits
the precision with which we may place bounds in the second particle sector, but there
should still be order of magnitude validity, assuming the LECs have natural sizes. This will make our
$\chi$PT results very powerful.

So the forms we have found for the effective Lagrangians in the pion and nucleon sectors have important
physical consequences. We shall first return to the pure pion sector and look more closely at the 
propagation terms. Written in terms of the physical fields, the free two-pion Lagrange density is
\begin{equation}
\label{lo_2pionphysical}
\mathcal{L}_{2\pi}^{\text{LO}}=\frac{\beta^{(1)}}{2}(c^{\mu\nu}_{u_{L}}+c^{\mu\nu}_{d_{L}}+
c^{\mu\nu}_{u_{R}}+c^{\mu\nu}_{d_{L}})(\partial^{\mu}\pi^{+}\partial^{\nu}\pi^{-} +
\partial^{\mu}\pi^{-}\partial^{\nu}\pi^{+} + \partial^{\mu}\pi^{0}\partial^{\nu}\pi^{0}).
\end{equation}
This takes a standard form for Lorentz violation involving a spin-0 field. In the
general Lagrange density
\begin{equation}
\label{spin0_general}
\mathcal{L}_{\text{spin-0}}=\frac{1}{2}\partial^{\mu}\phi_{a}\partial_{\mu}\phi_{a}
+\frac{1}{2}k^{\mu\nu}\partial_{\mu}\phi_{a}\partial_{\nu}\phi_{a}-\frac{m^{2}}{2}\phi_{a}\phi_{a},
\end{equation}
the tensor $k^{\mu\nu}$ modifies the equations of motion---or, equivalently, the energy-momentum
relation for free propagating particles. These propagation modifications can have many observable
physical consequences.

Making very precise laboratory measurements with short-lived particles such as pions can be very
challenging. As a result, most of the best constraints on Lorentz violation in the pion sector instead
come from high-energy astrophysical observations~\cite{ref-altschul14,ref-altschul16,ref-altschul35}.
When particles have modified energy-momentum relations,
which do not have the standard relativistic forms, there may be upper and lower thresholds for various
decay and emission processes. For example, with an appropriate choice of parameters, the decay of photons
into charged particle-antiparticle pairs (such as $\gamma\rightarrow\pi^{+}+\pi^{-}$) may occur for
sufficiently energetic $\gamma$-rays. Observations of TeV $\gamma$-rays that have traversed astrophysical
distances indicate that the threshold for this process, if it exists, must be above the energies of the
measured photons, which means that particular combinations of the $k_{\pi}^{\mu\nu}$ coefficients must
be correspondingly very small. A similar argument exists for another photon energy loss process that is ordinarily
forbidden, $\gamma\rightarrow\gamma+\pi^{0}$. Other important processes are $\pi^{0}\rightarrow\gamma+\gamma$,
the normal $\pi^{0}$ decay mode, which could become disallowed above a certain energy, or $\pi^{0}\rightarrow N+\bar{N}$,
which would instead become the dominant decay mode if it were energetically allowed, because of the larger
pion-nucleon coupling. Typically, astrophysical observations involving observed quanta at an energy $E$ allow
us to have constraints on combinations of $k_{\pi}^{\mu\nu}$ at the $\sim m_{\pi}^{2}/E^{2}$ level. In practice,
this means there are bounds at the $10^{-10}$--$10^{-13}$ levels, which are fairly strong. However, the
bounds are on complicated combinations of all the $k_{\pi}^{\mu\nu}$ coefficients, which are determined by the
sky coordinates of the sources involved. Moreover, there
are much stronger bounds in other sectors, and there are limited possibilities for improving the direct pion bounds,
since major improvements would require observations of substantially more energetic quanta, which can be few and
far between.

According to eqs.~\eqref{lo_2pionlagrangian} and~\eqref{lo_2pionphysical}
there is a single $k_{\pi}^{\mu\nu}$ tensor common to all the physical pion fields. The tensor takes the form
\begin{equation}
k_{\pi}^{\mu\nu}=\beta^{(1)}
(c_{u_{L}}^{\mu\nu}+c_{u_{R}}^{\mu\nu}+c_{d_{L}}^{\mu\nu}+c_{d_{R}}^{\mu\nu}).
\end{equation}
That the three pion types share
these same leading order Lorentz violation coefficients should be no surprise, since
in the chiral limit, the pion wave functions all
contain equal mixtures of the $u$ and $d$ fields, as well as equal right and left helicities.
At this point, we should recall that physical $SU(2)_{L}$ gauge invariance requires that the
underlying Lorentz violation coefficients for left-handed $u$ and $d$ quarks be the same,
$c_{u_{L}}^{\mu\nu}=c_{d_{L}}^{\mu\nu}=c_{q_{L}}^{\mu\nu}$. This relation would be of crucial
importance if we were seeking to relate the results of experiments performed on hadrons to
the fundamental quark coefficients.

However, we shall instead focus here on the more concrete problem of relating separate sets of
directly observable Lorentz violation coefficients. The key is that the combination
$c_{p}^{\mu\nu}+c_{n}^{\mu\nu}$ of readily measurable baryon parameters depends on the exact same
linear combination of quark parameters as the pion $k_{\pi}^{\mu\nu}$. The Lorentz-violating kinetic terms in
the effective Lagrangian for the nucleon sector of the SME are written in terms of
four coefficient tensors $c_{p}^{\mu\nu}$, $c_{n}^{\mu\nu}$, $d_{p}^{\mu\nu}$, and
$d_{n}^{\mu\nu}$. These enter the Lagrange density for a species of Dirac fermions as
\begin{equation}
\label{eq-Lspinhalf}
\mathcal{L}_{\mathrm{spin-}\frac{1}{2}}=\bar{\psi}\left[i(\gamma^{\mu}+c^{\nu\mu}\gamma_{\nu}+d^{\nu\mu}
\gamma_{5}\gamma_{\nu})D_{\mu}-m\right]\psi.
\end{equation}
(Additional dimension-three Lorentz-violating operators have been neglected.)
That $c_{p}^{\mu\nu}$, $c_{n}^{\mu\nu}$, $d_{p}^{\mu\nu}$, and $d_{n}^{\mu\nu}$
number four should be no surprise, since that is also
the number of independent tensors at the quark level, before $SU(2)_{L}$ gauge invariance is imposed.

In nonrelativistic experiments,
$c_{p}^{\mu\nu}$ and $c_{n}^{\mu\nu}$ receive contributions from the $\alpha^{(1)}$ and $\alpha^{(2)}$
terms, while $d_{p}^{\mu\nu}$ and $d_{n}^{\mu\nu}$ receive
contributions from the $\alpha^{(3)}$ and $\alpha^{(4)}$ terms.
Starting from~\eqref{eq-Lspinhalf}, the nonrelativistic Hamiltonian may be
determined using a relatively straightforward Foldy-Wouthuysen transformation~\cite{ref-foldy,ref-kost10},
and most measurements of Lorentz violation involving protons and neutrons
are done nonrelativistically, typically using atomic
clocks~\cite{ref-prestage,ref-lamoreaux,ref-chupp,ref-kost6,ref-wolf,ref-smiciklas}.
In this regime, it is possible to read off the effective coefficients directly from 
$\mathcal{L}_{\pi N}^{\text{LO}}$ with the pions neglected. For example, we obtain
\begin{equation}
\label{eq-cp}
c_{p}^{\mu\nu}=\left[\frac{1}{2}\alpha^{(1)}+\alpha^{(2)}\right] (c_{u_{L}}^{\mu\nu}+c_{u_{R}}^{\mu\nu})
+\left[-\frac{1}{2}\alpha^{(1)}+\alpha^{(2)}\right](c_{d_{L}}^{\mu\nu}+c_{d_{R}}^{\mu\nu}).
\end{equation}
One could attempt to use relations such as eq.~\eqref{eq-cp} to place disentangled bounds on
the $u$ and $d$ quark coefficients. However, the presence of the unknown LECs makes it
impossible to do this fully quantitatively. The results would be rather unsurprising
order of magnitude constraints on the quark-level parameters (which are not separately
observable anyway).

It would be possible to proceed a bit further using a ``quenched'' approximation, under which
the only quarks present in a nucleon are the valance quarks. In that case, the $u$ quark contribution
to the proton Lorentz violation should be twice the $d$ contribution, and we may infer that
$\alpha^{(1)}=\frac{2}{3}\alpha^{(2)}$. However, ignoring the presence of dynamical quark-antiquark pairs inside a
nucleon is obviously a drastic approximation, and it is not clear how much physical value the
$\alpha^{(1)}=\frac{2}{3}\alpha^{(2)}$ result has.

\begin{table}
\begin{center}
\begin{tabular}{|c|c|c|}
\hline
Coefficent & Proton Bound & Neutron Bound \\
\hline
$c_{Q}=c_{XX}+c_{YY}-2c_{ZZ}$ & $10^{-21}$ & $10^{-10}$ \\
$c_{-}=c_{XX}-c_{YY}$ & $10^{-24}$ & $10^{-28}$ \\
$c_{(XY)}$ & $10^{-24}$ & $10^{-29}$ \\
$c_{(XZ)}$ & $10^{-25}$ & $10^{-28}$ \\
$c_{(YZ)}$ & $10^{-25}$ & $10^{-28}$ \\
$c_{(TX)}$ & $10^{-20}$ & $10^{-5}$ \\
$c_{(TY)}$ & $10^{-20}$ & $10^{-5}$ \\
$c_{(TZ)}$ & $10^{-20}$ & $10^{-5}$ \\
$c_{TT}$ &  $10^{-11}$ & $10^{-11}$ \\
\hline
\end{tabular}
\caption{
\label{table-oldbounds}
Strengths of existing constraints on Lorentz violation in the proton and neutron sectors.
Symmetrized combinations are denoted $c_{(\mu\nu)}=c_{\mu\nu}+c_{\nu\mu}$.}
\end{center}
\end{table}

Summing eq.~\eqref{eq-cp} and the analogous formula for neutrons gives an expression that is
directly proportional to $(c_{u_{L}}^{\mu\nu}+c_{u_{R}}^{\mu\nu}+c_{d_{L}}^{\mu\nu}+c_{d_{R}}^{\mu\nu})$.
Again, this is not particularly surprising. The resulting coefficients are effectively averaged over
both spin and isospin, so they have equal contributions from all the quark tensors. Since this same
combination of quark coefficients occurs in the pion $k_{\pi}^{\mu\nu}$, it is possible to place
order-of-magnitude bounds on the $k_{\pi}^{\mu\nu}$ by combining observations made of the proton and neuton.
Better than order-of-magnitude accuracy is not possible, however, because of the presence of unknown LECs
in all the hadronic expressions, which at the moment can only be estimated using naive dimensional analysis.

Bounds on SME parameters are typically expressed in a system of Sun-centered celestial equatorial
coordinates, with the $Z$-axis pointing along rotation axis of the Earth, and the $X$-axis pointing to the
vernal equinox point on the celestial sphere. The $Y$-direction is determined by the right-hand rule,
and time in these coordinates is denoted by $T$. This coordinate system is well suited for describing the
results of many kinds of terrestrial experiments, particularly measurements of spatial anisotropy that
rely on the daily rotation of the Earth. Table~\ref{table-oldbounds} shows the best current order of
magnitude constraints for the $c_{p}^{\mu\nu}$ and $c_{n}^{\mu\nu}$ coefficients, although recent analyses based
on more careful nuclear models actually suggest significant improvements over some of these
constraints~\cite{ref-flambaum,ref-brown}.

\begin{table}
\begin{center}
\begin{tabular}{|c|c|}
\hline
Coefficient & Bound \\
\hline
$(k_{\pi})_{-}=(k_{\pi})_{XX}-(k_{\pi})_{YY}$ & $10^{-23}$ \\
$(k_{\pi})_{(XY)}$ & $10^{-23}$ \\
$(k_{\pi})_{(XZ)}$ & $10^{-24}$ \\
$(k_{\pi})_{(YZ)}$ & $10^{-24}$ \\
\hline
\end{tabular}
\caption{
\label{table-newbounds}
New constraints on pion Lorentz violation coming from comparisons to the nucleon sector.}
\end{center}
\end{table}

It is evident from table~\ref{table-oldbounds} that there are much stronger constraints in both the
proton and neutron sectors than for pions for four types of coefficients. Table~\ref{table-newbounds}
therefore quotes new bounds on four pion parameters. Since the LECs in the nucleon and meson sectors are
unknown but expected to be of $\mathcal{O}(1)$, we have set the pion constraints to be one order of magnitude weaker than the
looser of the
contributing proton and neutron bounds. Yet this still makes improvements of at least ten orders of
magnitude over direct astrophysical constraints on the same parameters.

\section{Conclusions and Outlook}
\label{sec-concl}

The ten order of magnitude improvement in pion sector constraints is evidence of how effective the
$\chi$PT method can be. We have produced disentangled bounds on four Lorentz violation coefficients
that affect pion propagation, without looking directly at any pions. These bounds are among our most
important results.

However, the effective Lagrange densities we have derived are also quite important. In the pure pion
sector, we have a systematic way of generating multi-pion vertices with definite relations between them imposed by chiral symmetry.
The lowest-order Lagrangian in the nucleon sector has also been
laid out, again showing new terms. In the future, this kind of analysis may lead to an understanding
of Lorentz violation for spin-1 and spin-$\frac{3}{2}$ composite particles, which have never really been
studied in any detail.

The hadronic terms we have discussed---the $k_{\pi}^{\mu\nu}$ for pions, and the $c^{\mu\nu}$ and
$d^{\mu\nu}$ for nucleons---are among the most important coefficients in the SME. These kinds of
terms typically grow in importance with increasing energy, leading to the kind of unconventional thresholds
discussed above. This is one reason why we chose to work with these kinds of terms.

However, there are still many more steps to be taken, and we have necessarily begun our $\chi$PT
analysis by considering only a subset of the mSME terms that are likely to affect hadrons.
The most fundamental omission has been that we have not considered any Lorentz violation in the
$SU(3)_{c}$ gauge sector. The strong interaction sector of the SME includes another set of pure
gauge interactions that could make a leading order contribution to the two-index hadronic tensors
such as $k_{\pi}^{\mu\nu}$ or $c_{p}^{\mu\nu}$. [There are also other possible Lorentz-violating terms
in the $SU(3)_{c}$ action which, for symmetry reasons, cannot contribute to the hadron coefficients
we have considered.]
A more complete analysis should include the effects of both gauge and quark Lorentz violation on
hadronic fields.

There are also other forms of Lorentz violation (many of which are, additionally, forms of CPT violation)
that may exist for composite hadrons. There are more quark-level operators, particularly those with mass
dimension 3, that will contribute in entirely different ways to symmetry violations by mesons and baryons.
Detailed consideration of these operators will be another important task for the future.

 Moreover, there may also be additional terms describing
possible Lorentz-violating interactions between hadrons and external fields. Although we did not consider
any external fields in this paper, it is obvious that there may be new interaction terms that look like
Lorentz-violating deformations of the conventional anomalous magnetic moment interaction.
Anomalous moments of baryons are typically large, and they often play a crucial role in atomic clock
experiments. So including their Lorentz-violating analogues may be equally crucial to understanding
how to interpret atomic clock tests of isotropy and boost invariance. Our ultimate understanding of how to use
$\chi$PT to analyze Lorentz-violating theories will certainly need to include consideration of these
questions regarding external fields.

The question of how to relate the underlying quark and gluon coefficients in the mSME to the coefficients
for composite hadrons has been one of the most important remaining puzzles in Lorentz-violating effective field
theory. We have looked at how quark-level operators translate into pion, proton, and neutron operators, using the
apparatus of $\chi$PT. This and other
recent work~\cite{ref-noordmans} demonstrate the power of the $\chi$PT technique. It
has enabled us to place new bounds (improved over previous ones by ten orders of magnitude) on pion-sector Lorentz violation,
without directly studying pions. And yet there is still much more remaining to be understood about the use of $\chi$PT in the
context of the SME.

\section*{Acknowledgments}
BA thanks J. P. Noordmans for helpful discussions.
This material is based upon work supported by the U.S. Department of Energy, Office of Science, Office of Nuclear Physics, under
Award Number DE-SC0010300 (RK and MRS).



\begin{thebibliography}{99}

\bibitem{ref-kost1}D. Colladay, V. A. Kosteleck\'{y}, Phys. Rev. D {\bf 55},
6760 (1997).
\bibitem{ref-kost2}D. Colladay, V. A. Kosteleck\'{y}, Phys. Rev. D {\bf 58},
116002 (1998).
\bibitem{ref-tables}V. A. Kosteleck\'{y}, N. Russell, Rev. Mod. Phys. {\bf 83},
11 (2011); updated as arXiv:0801.0287v8.
\bibitem{ref-weinberg}S. Weinberg, Physica A, {\bf 96}, 327 (1979).
\bibitem{ref-gasser1}J. Gasser, H. Leutwyler, Ann. Phys. (N.Y.) {\bf 158}, 142 (1984).
\bibitem{ref-gasser2}J. Gasser, H. Leutwyler, Nucl. Phys. B {\bf 250}, 465 (1985).
\bibitem{Scherer:2012xha}S.~Scherer, M.~R.~Schindler, {\em A Primer for Chiral Perturbation Theory}
(Springer, New York, 2012).
\bibitem{ref-noordmans}J. P. Noordmans, J. de Vries, R. G. E. Timmermans, 
Phys.\ Rev.\ C {\bf 94}, 025502 (2016).
\bibitem{ref-greenberg}O. W. Greenberg, Phys. Rev. Lett. {\bf 89}, 231602 (2002).
\bibitem{ref-colladay2}D. Colladay, P. McDonald, J. Math. Phys. {\bf 43}, 3554
(2002).
\bibitem{ref-altschul30}B. Altschul, Phys. Rev. D {\bf 86}, 045008 (2012).
\bibitem{ref-coleman1}S. Coleman, J. Wess, B. Zumino, Phys. Rev. {\bf 177}, 2239
(1969).
\bibitem{ref-callan1}C.~G.~Callan, Jr., S.~R.~Coleman, J.~Wess and B.~Zumino, Phys. Rev.  {\bf 177}, 2247 (1969).
\bibitem{ref-georgi}H.~Georgi, {\em Weak Interactions and Modern Particle Theory} (Benjamin/Cummings, Menlo Park, 1984).
\bibitem{Gasser:1987rb}J.~Gasser, M.~E.~Sainio and A.~Svarc, Nucl. Phys. B {\bf 307}, 779 (1988).
\bibitem{Manohar:1983md}A.~Manohar, H.~Georgi, Nucl. Phys. B {\bf 234}, 189 (1984).
\bibitem{ref-fettes}N. Fettes, U.-G. Meissner, M. Mojzis, S. Steininger, Annals Phys.
{\bf 283}, 273 (2000).
\bibitem{ref-altschul14}B. Altschul, Astropart. Phys. {\bf 28}, 380 (2007).
\bibitem{ref-altschul16}B. Altschul, Phys. Rev. D  {\bf 77}, 105018 (2008).
\bibitem{ref-altschul35}B. Altschul, Phys. Rev. D {\bf 93}, 105007 (2016).
\bibitem{ref-foldy}L. L. Foldy, S. A. Wouthuysen, Phys. Rev. {\bf 78}, 29
(1950).
\bibitem{ref-kost10}V. A. Kosteleck\'y and C. D. Lane, J. Math. Phys. {\bf 40},
6245 (1999).
\bibitem{ref-prestage}J. D. Prestage, J. J. Bollinger, Wayne M. Itano, D. J. Wineland,
Phys. Rev. Lett. {\bf 54}, 2387 (1985).
\bibitem{ref-lamoreaux}S. K. Lamoreaux, J. P. Jacobs, B. R. Heckel, F. J. Raab,
E. N. Fortson, Phys. Rev. Lett. {\bf 57}, 3125 (1986).
\bibitem{ref-chupp}T. E. Chupp, R. J. Hoare, R. A. Loveman, E. R. Oteiza, J. M.
Richardson, M. E. Wagshul, A. K. Thompson, Phys. Rev. Lett. {\bf 63}, 1541 (1989).
\bibitem{ref-kost6}V. A. Kosteleck\'{y}, C. D. Lane, Phys. Rev. D {\bf 60},
116010 (1999).
\bibitem{ref-wolf}P. Wolf, F. Chapelet, S. Bize, A. Clairon,  Phys. Rev. Lett.
{\bf 96}, 060801 (2006).
\bibitem{ref-smiciklas}M. Smiciklas, J. M. Brown, L. W. Cheuk, S. J. Smullin,
M. V. Romalis, Phys. Rev. Lett. {\bf 107}, 171604 (2011).
\bibitem{ref-flambaum}V. V. Flambaum, Phys.\ Rev.\ Lett.\  {\bf 117}, 072501 (2016).
\bibitem{ref-brown}B. A. Brown, G. F. Bertsch, L. M. Robledo, M. V. Romalis, V.
Zelevinsky, arXiv:1604.08187.

\end{thebibliography}
\end{document}